\documentclass[journal=jctcce,manuscript=article]{achemso}
\setkeys{acs}{articletitle=true}

\usepackage{amsmath}
\usepackage{booktabs} 
\usepackage{caption} 
\usepackage{subcaption} 
\usepackage{graphicx}
\usepackage{pgfplots}
\usepackage{xr}

\usepackage[all]{nowidow}
\usepackage[utf8]{inputenc}
\usepackage{tikz}
\usetikzlibrary{er,positioning,bayesnet}
\usepackage{multicol}
\usepackage{algpseudocode,algorithm,algorithmicx}
\usepackage{hyperref}
\usepackage{cleveref}
\usepackage[inline]{enumitem} 

\definecolor{blue}{HTML}{1F77B4}
\definecolor{orange}{HTML}{FF7F0E}
\definecolor{green}{HTML}{2CA02C}

\pgfplotsset{compat=1.14}

\title[]{Boltz-ABFE: Free Energy Perturbation without Crystal Structures}

\author{Stephan Thaler}
\altaffiliation{These authors contributed equally.}
\affiliation{Valence Labs, 6666 Rue Saint-Urbain 100, Montréal, QC H2S 3H1, Canada}
\alsoaffiliation{Recursion Pharmaceuticals Inc, Salt Lake City, UT 84101, USA}

\author{Zhiyi Wu}
\altaffiliation{These authors contributed equally.}
\affiliation{Recursion Pharmaceuticals Inc, Salt Lake City, UT 84101, USA}

\author{William G. Glass}
\affiliation{Recursion Pharmaceuticals Inc, Salt Lake City, UT 84101, USA}

\author{Richard T. Bradshaw}
\affiliation{Recursion Pharmaceuticals Inc, Salt Lake City, UT 84101, USA}

\author{Gail Bartlett}
\affiliation{Recursion Pharmaceuticals Inc, Salt Lake City, UT 84101, USA}

\author{Prudencio Tossou}
\affiliation{Valence Labs, 6666 Rue Saint-Urbain 100, Montréal, QC H2S 3H1, Canada}
\alsoaffiliation{Recursion Pharmaceuticals Inc, Salt Lake City, UT 84101, USA}

\author{Geoffrey P. F. Wood}
\email{geoffrey.wood@recursion.com}
\affiliation{Recursion Pharmaceuticals Inc, Salt Lake City, UT 84101, USA}

\begin{document}
\maketitle              
\begin{abstract}
Free energy perturbation (FEP) is considered the gold-standard simulation method for estimating small molecule binding affinity, a quantity of vital importance to drug discovery. 
The accuracy of FEP critically depends on an accurate model of the protein-ligand complex as an initial condition for the underlying molecular dynamics simulation. This requirement has limited the impact of FEP in earlier stages of the discovery process, where appropriate experimental crystal structures are rarely available.
The latest generation of structure prediction models, such as Boltz-2, promise to overcome this limitation by predicting protein-ligand complex structures. In this work, we combine Boltz-2 with our own absolute FEP protocol to build Boltz-ABFE, a robust pipeline for estimating the absolute binding free energies (ABFE) in the absence of experimental crystal structures.
We investigate the quality of the structures predicted by Boltz-2, propose automated approaches to improve structures for use in molecular dynamics simulations, and demonstrate the effectiveness of the Boltz-ABFE pipeline for four protein targets from the FEP+ benchmark set.
Demonstrating the feasibility of absolute FEP simulations without experimental crystal structures, Boltz-ABFE significantly expands the domain of applicability of FEP, paving the way towards accelerated early-stage drug discovery via accurate, structure-based affinity estimation.

\end{abstract}
\section{Introduction}

The estimation of small molecule binding affinity has been a long standing, and vital, application of biomolecular simulation in drug discovery. To this end, the workhorse of affinity predictions for prospective drug candidates has been alchemical Relative and Absolute Binding Free Energy Perturbation (R/A-BFE) calculations using classical force fields. Since the first calculations of this type occurred in the 1980s\cite{jorgensen1985monte,wong1986dynamics,kollman1993free,wang2015accurate,michel2010prediction} significant advancements in algorithms,\cite{abel2017advancing} force-field development procedures\cite{jorgensen1988opls,jorgensen1996development,damm1997opls,mcdonald1998development,riniker2012developing,bayly1993well,cornell1995second,brooks2009charmm,vanommeslaeghe2010charmm,van2002gromos,oostenbrink2004biomolecular,hwang1994derivation,maple1994derivation,sun1998compass,ponder2003force,ponder2010current,weiner1986all,tian2019ff19sb,huang2017charmm36m,lu2021opls4,xue2023development,robertson2015improved,mobley2018escaping,boothroyd2023development,qiu2021development,wang2020end,wang2024open}and computer hardware\cite{pandey2022transformational} have enabled these calculations to be applied to hundreds of ligands at a time with accuracies in broad datasets approaching the 1 kcal/mol range\cite{Ross_fepplus2023, Hahn_openffbench2022}.

There are several factors that influence how well free energy calculations correlate with experimental affinities ($IC_{50}$, $K_{i}$ and $K_d$), generally falling into two main categories: First, those related to the simulation protocol (e.g. force-field, simulation length, $\lambda$-schedule, enhanced sampling methods), and second, those stemming from structure preparation (e.g. ligand pose, loop building, tautomer and charge state determination). In traditional target-based drug design, structure preparation is typically initiated from an experimentally determined X-ray or cryo-electron microscopy (Cryo-EM) structure. However, uncertainties intrinsic to the experimental methods can also lead to uncertain choices in preparation. For example, parts of the protein may be unresolved or missing, ligand or sidechain charge and tautomer states may be incorrect or ambiguous, and co-factors and other binding partners might be absent.  Although some preparation software packages and algorithms \cite{noauthor_spruce_nodate} aim to address many of these issues in a semi-automated manner, significant manual intervention and human expertise is often required in structure preparation.

An additional challenge arises from the unknown similarity of protein structures and ligand poses when bound to different ligand chemistry. During lead optimization, the number of available crystal structures is often low, as it is impractical to resolve them for every compound of interest. Instead, a single reference structure is commonly used as a template for docking algorithms, serving as the foundation for all subsequent 3D property predictions. This limits the range of applicability to compounds that are similar to the co-crystallized ligand to ensure the ligand binding pose is accurately captured, which is essential for accurate R/A-BFE results \cite{schindler2020large}.
In the early stages of drug discovery, such as hit identification and hit-to-lead, the situation is even more challenging, as even single reference crystal structures are unavailable or the target may not yet be identified at all. This renders 3D prediction of properties such as potency infeasible. 

Given the significant potential of FEP calculations to enhance early stages of the drug discovery process\cite{feng2022absolute, chen2023enhancing}, new approaches have been used to tackle the issue of protein-ligand complex prediction from different angles. First, machine learning-based docking methods \cite{corso2023diffdock, pei2023fabind, zhou2023uni, corso2024deep} attempt to more reliably predict the poses of the ligand in a target-agnostic way, but still require careful preparation or validation of the protein structure bound to the predicted pose. Moreover, while these ML docking methods often perform well as measured by root-mean-squared distance (RMSD), traditional docking methods are more robust and reliable in predicting chemically valid ligand poses \cite{buttenschoen2024posebusters}.
    
Alternatively, predicted protein apo structures, such as those predicted by AlphaFold2 \cite{jumper2021highly}, can be repurposed for docking calculations to avoid the need for experimental crystal structures. However, obtaining accurate ligand poses based on a predicted apo structure is a challenge \cite{beuming2022deep, coskun2023using}. In contrast, AlphaFold3 \cite{abramson2024accurate}, and open-source counterparts, such as Boltz-1/2 \cite{wohlwend2024boltz,passaro2025boltz}, can predict the protein holo structure co-folded with the ligand, overcoming the main limitations of both AlphaFold2 and ligand-only pose prediction methods. Additionally, AlphaFold3 and Boltz-2 predict confidence scores for the protein-ligand binding pose, and it is natural to start making assumptions that confidence scores might be related to measures of affinity. However, they were not found to correlate well \cite{zheng2025alphafold3}. Thus, a hybrid approach that uses predicted protein-ligand complexes as starting points for physics-based free energy calculation offers an intriguing alternative to traditional approaches. To this end, Scheen et al. \cite{scheen2025leveraging} recently used DragonFold and OpenFE \cite{openfe} to calculate relative binding free energies on the public target PFKFB3 and a proprietary structure (CHARM Therapeutics), with encouraging results. 

In this work, we explore this idea further by using the Boltz-1 and Boltz-2 models\cite{wohlwend2024boltz} for co-folding predictions on a more extensive set of targets that include structures the models have not seen before. We combine this with our own ABFE protocol\cite{wu2025optimizing} and demonstrate that Boltz predicted protein-ligand complexes can be used to initialize simulations that accurately estimate the free energy of binding ($\Delta$G), provided that some care is taken when choosing which structure prediction is taken forward for use in MD simulations. To exemplify the approach and the necessary choices, we present a robust pipeline that prepares Boltz-predicted structures for MD by automating the removal of common defects in the predicted structures such as overlapping atoms and incorrect ligand stereochemistry. We highlight common pitfalls for accurate structure prediction, employ the Boltz-1/2 models in a cross-docking study emulating target deconvolution applications and showcase the effectiveness of our 'Boltz-ABFE' pipeline using 4 protein targets from the FEP+ benchmark set\cite{Ross_fepplus2023, wang2015accurate}.

\

\section{Results}

\subsection{Boltz-ABFE}

\subsubsection{Chemically accurate ligand pose prediction via a coupled co-folding and re-docking approach}

\begin{figure}
    \centering
    \includegraphics[width=\linewidth]{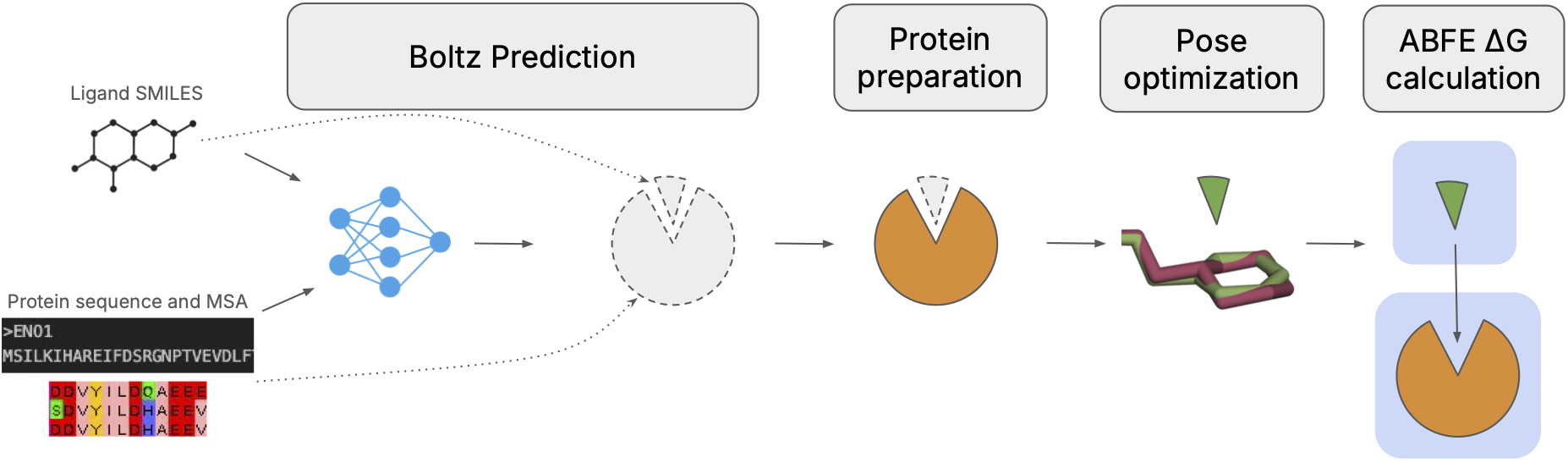}
    \caption{The Boltz-ABFE pipeline uses Boltz-1/Boltz-2 to predict the protein-ligand complex given the SMILES of the ligand and sequence(s) of the protein chain(s). Afterwards, the predicted protein receptor is further prepared for docking and molecular dynamics applications. To correct ligand chemistry errors, the known ligand is re-docked into the receptor using the predicted complex as a template. The resulting structure is then relaxed and equilibrated before performing the ABFE simulation.}
    \label{fig:Boltz-1-ABFE}
\end{figure}

The goal of Boltz-ABFE is to accurately predict the protein-ligand binding affinity from the compound's SMILES string and protein sequence information alone (Figure \ref{fig:Boltz-1-ABFE}). To this end, we used the recently introduced Boltz-1 and Boltz-2 models\cite{wohlwend2024boltz} to predict the structure of the protein-ligand complex from the sequence data. 

\begin{figure}
    \centering
    \includegraphics[width=1.0\linewidth]{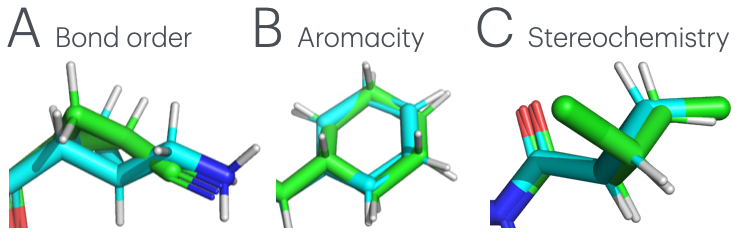}
    \caption{Common chemical inaccuracies in Boltz-1-generated ligand structures and the effects of refinement strategies. Boltz-1 occasionally models the input SMILES incorrectly, leading to incorrect bond orders (A), improper aromaticity (B), or incorrect stereochemistry (C).}
    \label{fig:artifacts}
\end{figure}

The basic Boltz-1 model, which predicts biomolecular complexes in the gas-phase without explicit hydrogen atoms, can generate structures with inaccuracies that are unsuitable for direct use in molecular dynamics (MD) simulations. These artifacts commonly manifest as steric clashes between protein side chains or as errors in ligand modeling, including incorrect bond orders (Figure \ref{fig:artifacts}A), aromaticity (Figure \ref{fig:artifacts}B), and stereochemistry (Figure \ref{fig:artifacts}C). When tested on four sets of target from the FEP+ benchmark \cite{Ross_fepplus2023} (TYK2, CDK2, JNK1, and P38), we observed that the four sets exhibited errors in these categories (Table \ref{tab:chemistry}).

To improve the physical quality of generated structures, optional inference-time steering potentials to guide the diffusion process were recently introduced to the Boltz models (e.g. Boltz-1x). We included these steering potentials in our testing of the Boltz-2 model and saw significant improvements in correcting for bond orders and aromaticity, but resolving incorrect stereochemistry remained a challenge (Table \ref{tab:chemistry}). This last type of error remains a detrimental one for successful MD simulations, as incorrect stereoisomers often present themselves as activity-cliffs, which cannot be corrected by standard energy minimization. To address these critical errors in ligand chemistry, we designed a re-docking protocol. In this procedure, the initial ligand is docked back into the co-folded receptor using traditional docking software, with the predicted complex serving as a template for geometry and interactions.\\

\begin{table}[h!]
\centering
\caption{Comparison of failure rates between Boltz-1 and Boltz-2 models across different error categories for four protein targets. Values represent the percentage of models exhibiting the specified error.}
\label{tab:chemistry}
\begin{tabular}{lcccccccc}
\toprule
& \multicolumn{2}{c}{Steric Clash} & \multicolumn{2}{c}{Aromaticity} & \multicolumn{2}{c}{Bond Order} & \multicolumn{2}{c}{Stereochemistry} \\
\cmidrule(lr){2-3} \cmidrule(lr){4-5} \cmidrule(lr){6-7} \cmidrule(lr){8-9}
\textbf{Protein} & \textbf{Boltz-1} & \textbf{Boltz-2} & \textbf{Boltz-1} & \textbf{Boltz-2} & \textbf{Boltz-1} & \textbf{Boltz-2} & \textbf{Boltz-1} & \textbf{Boltz-2} \\
\midrule
TYK2   &  0.00 & 0.00 &   0.00 & 0.00 & 6.88 & 0.00 & 23.75 & 15.63 \\
CDK2   &  0.00 & 0.00 & 100.00 & 0.00 & 0.00 & 0.00 &  0.00 & 4.38 \\
JNK1   & 10.48 & 0.00 &   0.00 & 0.00 & 0.00 & 0.00 &  N/A & N/A \\
P38    &  0.29 & 0.00 &   4.12 & 0.00 & 7.35 & 0.00 &  5.39 & 6.76 \\
TOTAL  &  2.64 & 0.00 &  20.00 & 0.00 & 4.14 & 0.00 &  6.55 & 6.32 \\
\bottomrule
\end{tabular}

\end{table}

We compared the redocking approach using two template-based OpenEye docking methods: POSIT and Hybrid\cite{posit2015,fredhybrid2012}. For ligands with high fingerprint similarity to the template, POSIT first aligns the target (initial) and template (Boltz) poses with a Maximum Common Substructure (MCS) overlay, followed by a shape- and pharmacophore-based fitting and optimization. Therefore poses generated by POSIT faithfully recapitulate the template position and geometry, but a key limitation arises when the Boltz prediction contains incorrect stereochemistry. In such cases, POSIT's emphasis on rigid shape preservation can generate highly strained, unfavorable conformers. To circumvent this potential issue, we also evaluated the Hybrid docking protocol. The Hybrid approach scores potential ligand poses using information from both the protein and template ligand, but does not adjust the internal ligand geometry to fit the template structure. Finally, we performed absolute binding free energy (ABFE) calculations on predicted structures of TYK2 complexes\cite{wang2015accurate} to evaluate the suitability of each approach for Boltz structure preparation. 

As shown in the SI Table S1, POSIT achieved the best agreement between calculated and experimental binding free energies, whereas the Hybrid approach was less accurate. Visual inspection indicated that Hybrid docking could yield suboptimal ring placement. This potential limitation arose because the protocol relies on docking pre-generated conformers; if the conformer necessary for optimal placement (particularly of rings) is absent from the initial ensemble, it cannot be recovered during the subsequent docking stage. Based on these findings, we proceeded with OpenEye POSIT for ligand chemistry correction. Figure S1 presents some of the differences in chemistry for the TYK2 ligands and the proposed poses of the corrected ligands generated by the two re-docking approaches

To address steric clashes in predicted structures, which hinder processing through the OpenEye Spruce preparation pipeline, we adopted the strategy of generating multiple models for each protein-ligand pair. This approach leverages the probability that such clashes will not be present in every generated pose. We used this multi-model strategy to compare the performance of Boltz-1 and Boltz-2, generating 10 models per complex and assessing the proportion that could be successfully processed.

\begin{figure}
    \centering
    \includegraphics[width=1.0\linewidth]{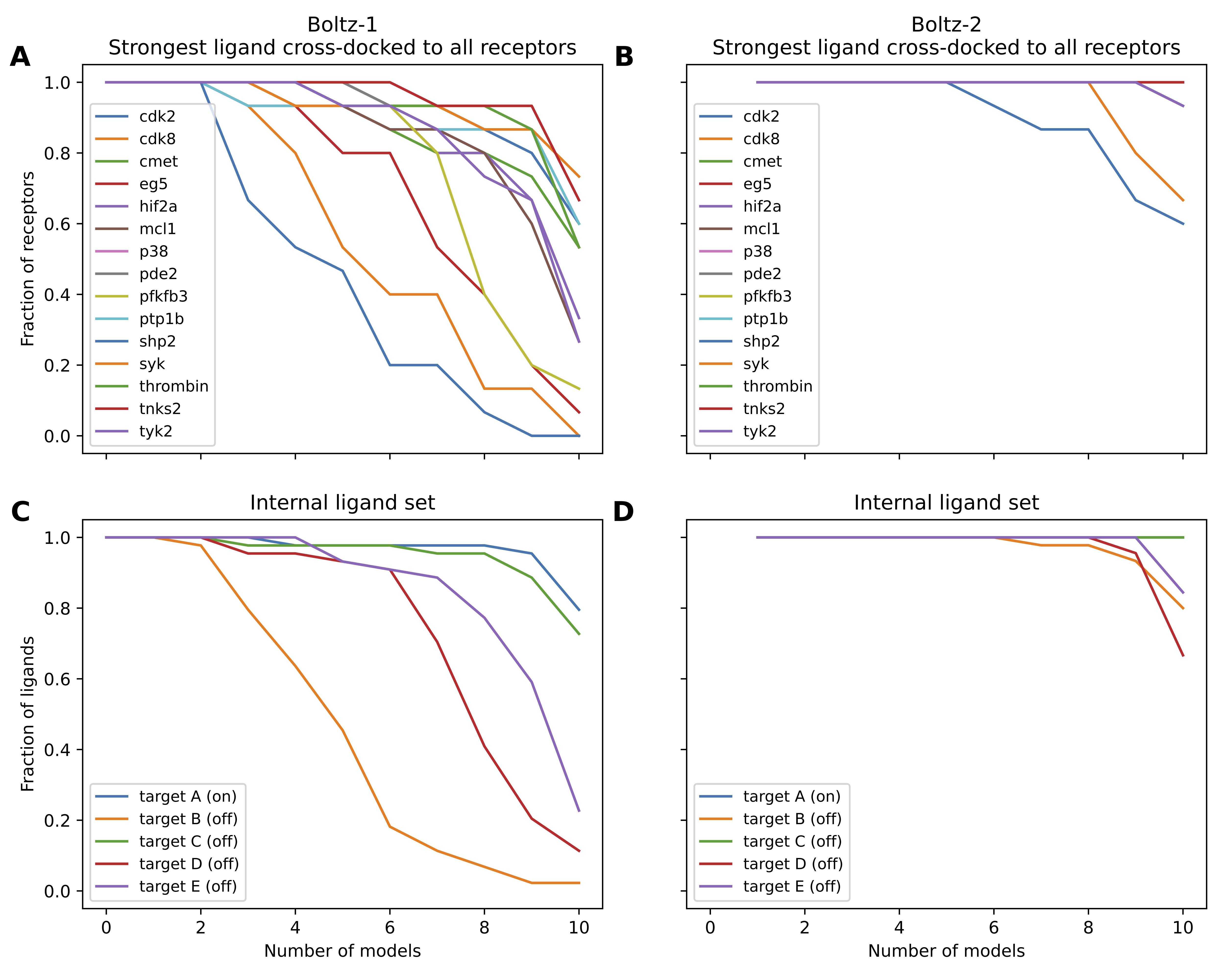}
    \caption{Comparison of success rates for the Boltz-1 and Boltz-2 models in generating clash-free structures across two co-folding scenarios. The y-axis represents the percentage of ligands yielding at least X successful models (x-axis). The figure compares the performance of Boltz-1 and Boltz-2 on two tasks: (i) cross-docking the highest-affinity ligand from each of the 15 OpenFF benchmark systems to the remaining targets (A: Boltz-1, B: Boltz-2), and (ii) docking internal ligand sets to their on-targets and related off-targets (C: Boltz-1, D: Boltz-2).}
    \label{fig:success_rate}
\end{figure}

Our evaluation revealed that the Boltz-2 model yields structures with significantly fewer steric clashes than Boltz-1. While detailed results across several scenarios are available in the Supporting Information (see Figure S2 for Boltz-1 and Figure S3 for Boltz-2), we highlight two cases here where Boltz-2 has a markedly higher success rate (Figure \ref{fig:success_rate}). These cases are: (1) cross-docking the strongest binder from each of the 15 OpenFF benchmark systems to all 15 targets, and (2) docking an internal ligand set to the known on-target compared with related off-targets. Although generating 10 models per pair generally proved sufficient to yield at least one clash-free structure, Boltz-2 is notably more efficient and robust in these challenging co-folding and re-docking scenarios.

\subsubsection{Classical scoring functions may aid model selection but are insufficient for target deconvolution}

Given that Boltz-2 successfully co-folded ligands against multiple targets, we explored the possibility that co-folded structures and rapid classical scoring functions could be used in a target deconvolution approach. Specifically, we wanted to know if this process could detect the native on-target from a set of potential targets without further need for free energy simulations. To this end, we scored cross-docked poses from the OpenFF benchmark systems (Figure \ref{fig:success_rate}B) and internal benchmark systems (Figure \ref{fig:success_rate}D) with the classical docking score Chemgauss4 and a semi-empirical QM protein-ligand interaction energy score calculated via a Fragment Molecular Orbital (FMO) approach \cite{nishimoto2016fragment, guareschi2023sophosqm}. For the OpenFF benchmark systems there is no available cross-target binding data, but we assumed there was no appreciable cross-target affinity for each ligand. For the internal set, all ligands showed higher affinity for the on-target, but some non-negligible affinity for off-targets. This might be related to the high sequence and structural similarity between on- and off-targets for the internal set, making them a more challenging dataset for such a rapid deconvolution approach.

\begin{figure}
    \centering
    \includegraphics[width=1.0\linewidth]{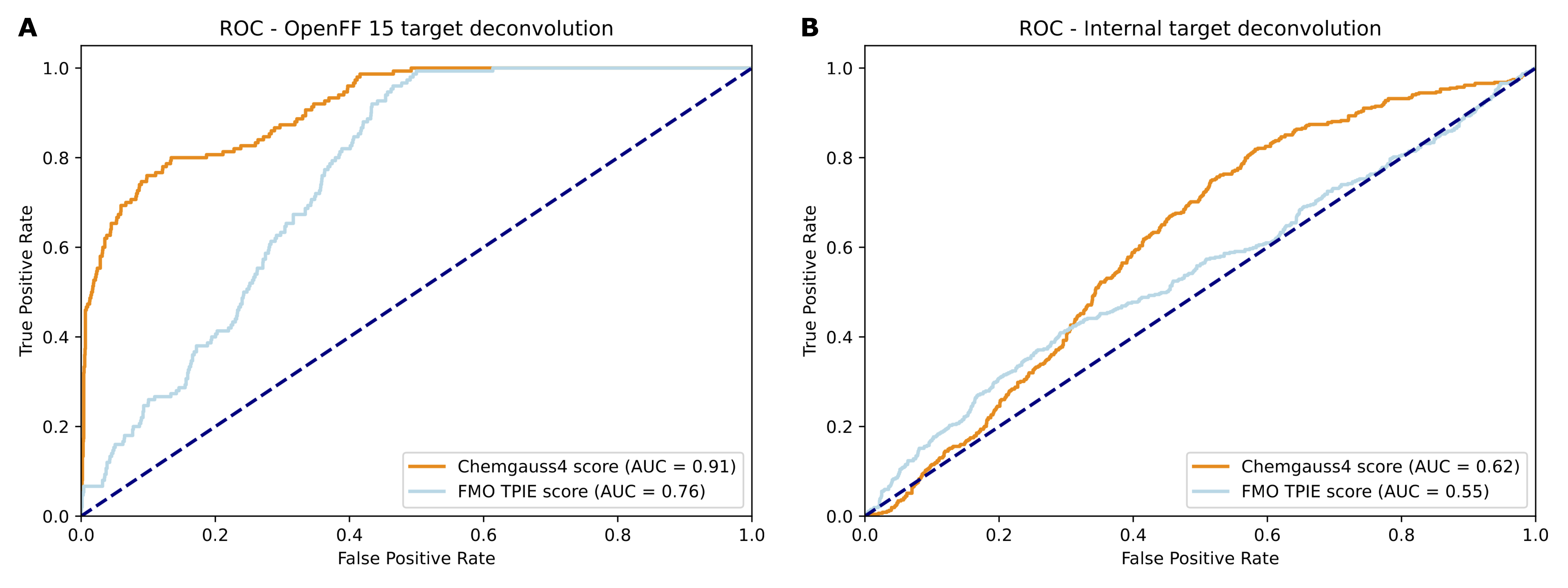}
    \caption{Receiver Operator Characteristic (ROC) plots for correct target deconvolution of cross-docked ligand poses, scored with either Chemgauss4 (orange) or the Total Protein Interaction Energy (TPIE) of the semi-empirical FMO approach (blue). Area under the curve (AUC) values are calculated for each scoring function (A) Highest affinity ligand of each OpenFF target cross-docked to all receptors, (B) Internal ligand set of 1 on-target and 5 off-targets.}
    \label{fig:deconvolution_roc}
\end{figure}

Target deconvolution with simple scorers was highly successful for the most potent ligands from the OpenFF set (Figure \ref{fig:deconvolution_roc}A), using either the Chemgauss4 score (AUC = 0.91) or the FMO TPIE score (AUC = 0.76). However, AUC values for the more challenging internal target dataset were much lower, AUC = 0.62 using Chemgauss4 as the scorer, or AUC = 0.55 using the FMO TPIE (Figure \ref{fig:deconvolution_roc}B).

These results suggest that in cases like the OpenFF benchmark set, where the target protein shares little similarity in sequence, structure, or ligand affinity to potential off-targets, simple classical scorers can be an efficient filter of incorrect targets. However, for targets with similar structures and relative ligand affinities, like the internal set, classical scorers cannot sufficiently identify the native target or rank selectivity.

\subsubsection{Generating reliable co-folded structures for MD simulations requires careful tailoring of sequence and biological assembly}

\begin{figure}
    \centering
    \includegraphics[width=1.0\linewidth]{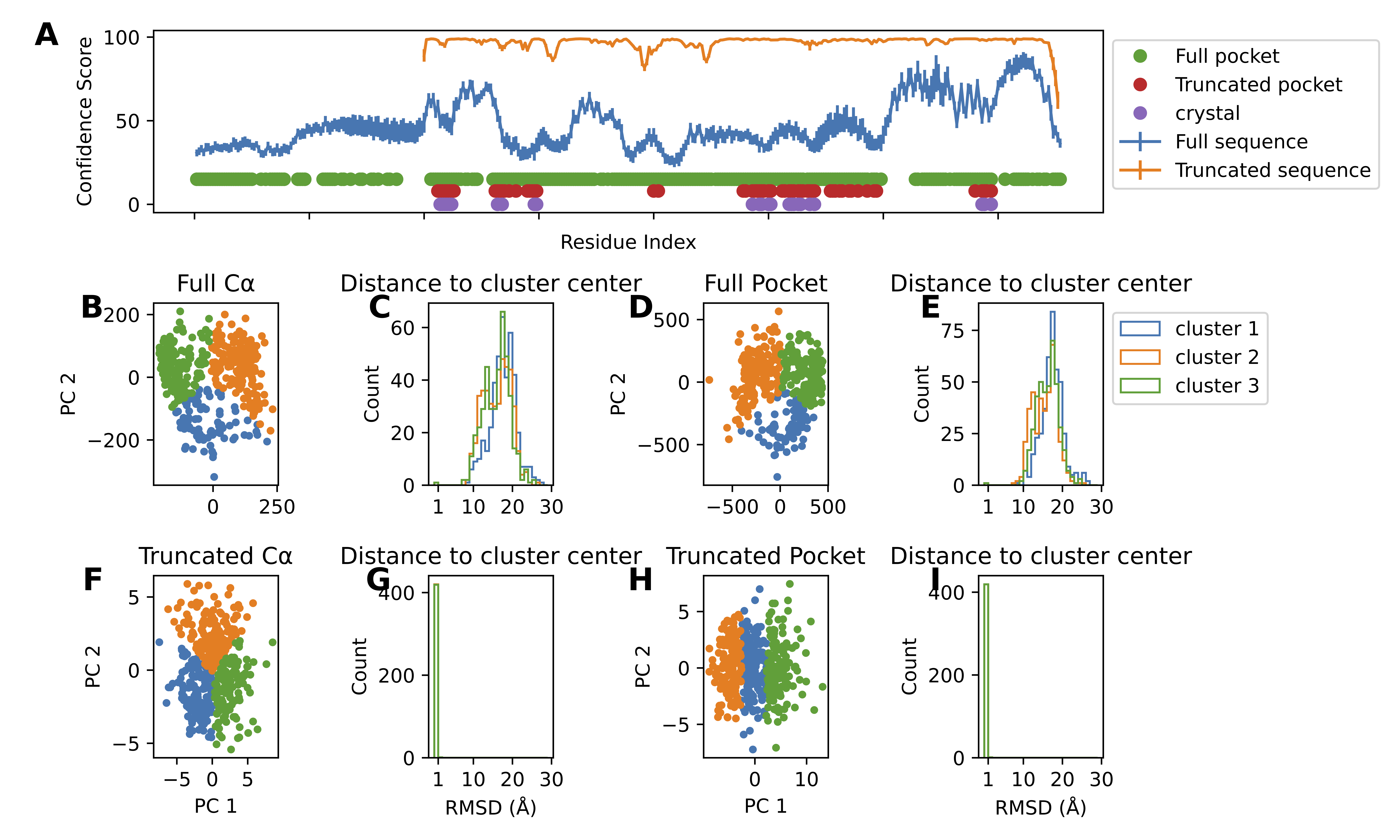}
    \caption{Truncating low-confidence regions improves model confidence scores and structural validity. (A) Per-residue confidence scores for models generated using the full sequence (blue) and truncated sequence (orange). Residues within 6 Å of the ligand are highlighted for the full-sequence model (green), truncated-sequence model (red) and crystal homolog (purple). (B, F) Principal component analysis (PCA) of C$\alpha$ atom coordinates, projected into two dimensions and clustered into three groups for models generated from the full sequence (B) and truncated sequence (F). (C, G) Root-mean-square deviation (RMSD) distributions of each model relative to the centers of the three clusters for the full-sequence (C) and truncated-sequence (G) models. (D, H) PCA analysis based on heavy atoms in the binding pocket, defined as residues within 6 Å of the ligand, for models generated from the full sequence (D) and truncated sequence (H). (E, I) RMSD distributions relative to the three cluster centers for the heavy atoms in the binding pocket for the full-sequence (E) and truncated-sequence (I) models.}
    \label{fig:truncation}
\end{figure}

In the absence of prior knowledge, the entire UniProt protein sequence can be used to generate Boltz-1/2 models. However, for targets that contain intrinsically unstructured regions or that are sparsely represented in the PDB, this often yields models with extensive low-confidence regions, especially at the termini, sometimes exceeding 100 residues (Figure \ref{fig:truncation}A). Visual inspection of these regions indicates potentially unrealistic secondary structures. A related concern is that modeled ligands may form extensive interactions with these low-confidence regions, which are not observed in homologous crystal structures, suggesting they are spurious artifacts. Furthermore, these regions introduce significant structural heterogeneity within the set of generated models. A principal component analysis (PCA) using the C$\alpha$ atoms demonstrates this variability (Figure \ref{fig:truncation}B), with root-mean-square deviations (RMSDs) to cluster centers often surpassing 20 Å (Figure \ref{fig:truncation}C).\\

 High structural variance is also evident among binding pocket heavy atoms (within 6 Å of the ligand) (Figure \ref{fig:truncation}D, E). Such conformational uncertainty complicates binding free energy calculations, as low confidence region encroachment into the binding site can produce inaccurate and unstable pocket definitions. To address this, we truncated predicted low-confidence regions before generating Boltz-1/2 models. This approach enhanced model confidence scores, particularly for binding pocket residues, and substantially decreased structural variability, both for global C$\alpha$ atoms (Figure \ref{fig:truncation}G, G) and binding site heavy atoms (Figure \ref{fig:truncation}H, I)\\
 
\begin{figure}
    \centering
    \includegraphics[width=1.0\linewidth]{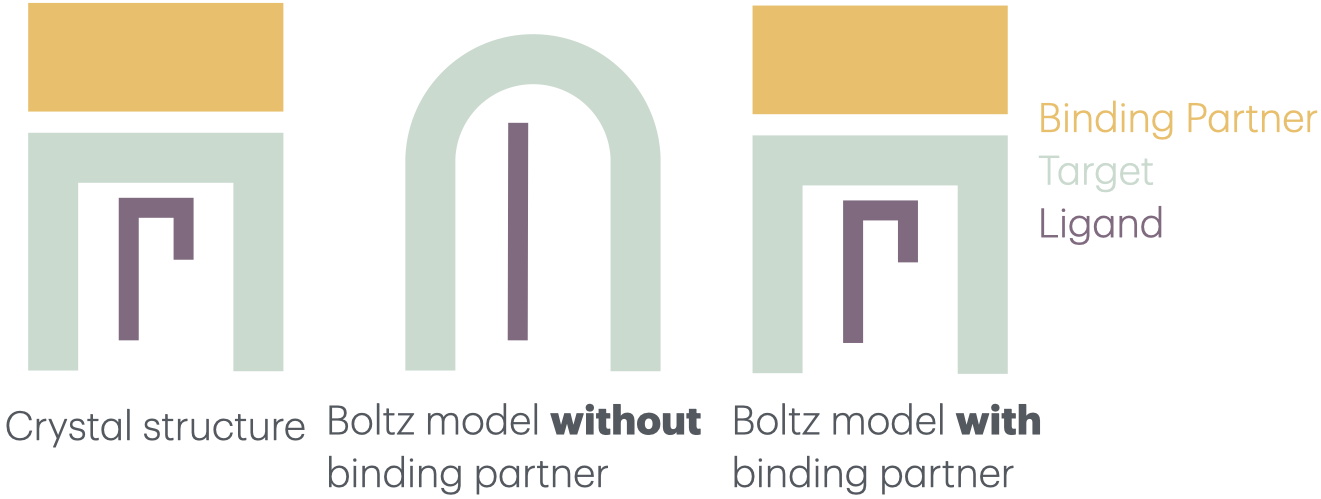}
    \caption{Schematic representation of ligand binding conformations in the presence and absence of a binding partner. In the crystal structure, the ligand adopts a folded conformation within the binding pocket. When Boltz-2 models are generated without the binding partner, the ligand is predicted in an extended conformation, with the binding pocket artificially expanding into the space normally occupied by the binding partner. Including the binding partner during Boltz-2 model generation restores the ligand’s folded conformation and the native binding pocket, aligning more closely with the crystal structure.}
    \label{fig:binding_partner}
\end{figure}
\begin{figure}
    \centering
    \includegraphics[width=1.0\linewidth]{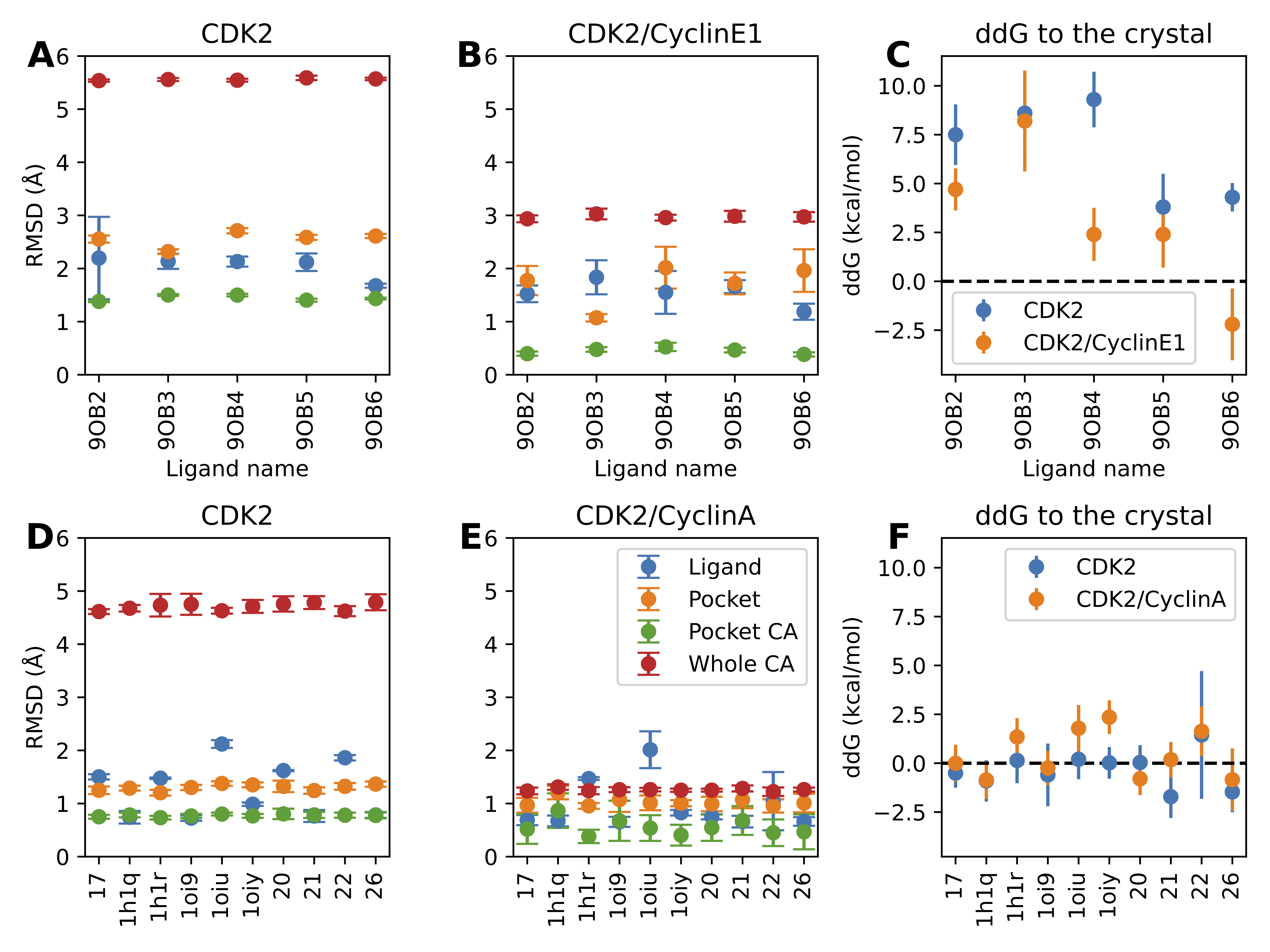}
    \caption{(A) RMSD of ligand heavy atoms, pocket heavy atoms, pocket C$\alpha$ atoms, and all C$\alpha$ atoms between Boltz-2-generated CDK2 models and the CDK2/CyclinE1 crystal structures (PDB: 9OB2-9OB6). (B) RMSD between CDK2 in the Boltz-2 model of the CDK2/CyclinE1 complex and CDK2 in the corresponding crystal structure. (C) $\Delta\Delta$G between ABFE $\Delta$G computed on the crystal structure and on the Boltz-2 model with CDK2 alone or in complex with CyclinE1. (D) RMSD between CDK2 in the Boltz-2 model and the CDK2 crystal structure. (E) RMSD between CDK2 in the CDK2/CyclinA Boltz-2 model and the CDK2 crystal structure. (F)  (H) $\Delta\Delta$G between ABFE $\Delta$G computed on the crystal structure and on the Boltz-2 model with CDK2 alone or in complex with CyclinA.}
    \label{fig:binding_partner_ddG}
\end{figure}

Beyond challenges identifiable from sequence and ligand data alone, accurate modeling sometimes requires additional biological context. A prime example is proteins that need binding partners to adopt their native conformation, such as cyclin-dependent kinases (CDKs). When co-folding a ligand with CDK alone, particularly for larger ligands ($>$30 heavy atoms) underrepresented in the training data, the model can predict an artificially expanded binding pocket and an incorrect, extended ligand conformation. However, including the cyclin binding partner restores the pocket to its correct shape and induces the proper folded ligand conformation, consistent with crystal structures (Figure \ref{fig:binding_partner}). We demonstrated this using five recently deposited PDB structures (9OB2-9OB6). Boltz-2 models generated without the binding partner showed significant deviations in global C$\alpha$ positions, pocket heavy atoms, and ligand poses compared to the crystal structures (Figure \ref{fig:binding_partner_ddG}A). Including the partner restored these features to biologically relevant conformations (Figure \ref{fig:binding_partner_ddG}B), which in turn improved absolute binding free energy (ABFE) predictions, bringing them closer to results derived from crystal structures (Figure \ref{fig:binding_partner_ddG}C).

In contrast, this effect was less pronounced for the older CDK2 benchmark set. While including the binding partner resulted in a moderate improvement in C$\alpha$ positioning, pocket atom alignment, and ligand pose (Figure \ref{fig:binding_partner_ddG}D and E), subsequent ABFE calculations did not show a significant improvement in binding free energy predictions (Figure \ref{fig:binding_partner_ddG}F). These findings suggest that for protein-ligand systems not well-represented in the training set, incorporating key binding partners is a valuable strategy for refining Boltz-2 predictions and improving downstream calculations.

\subsection{FEP+ benchmark}
\begin{figure}
    \centering
    \includegraphics[width=1.0\linewidth]{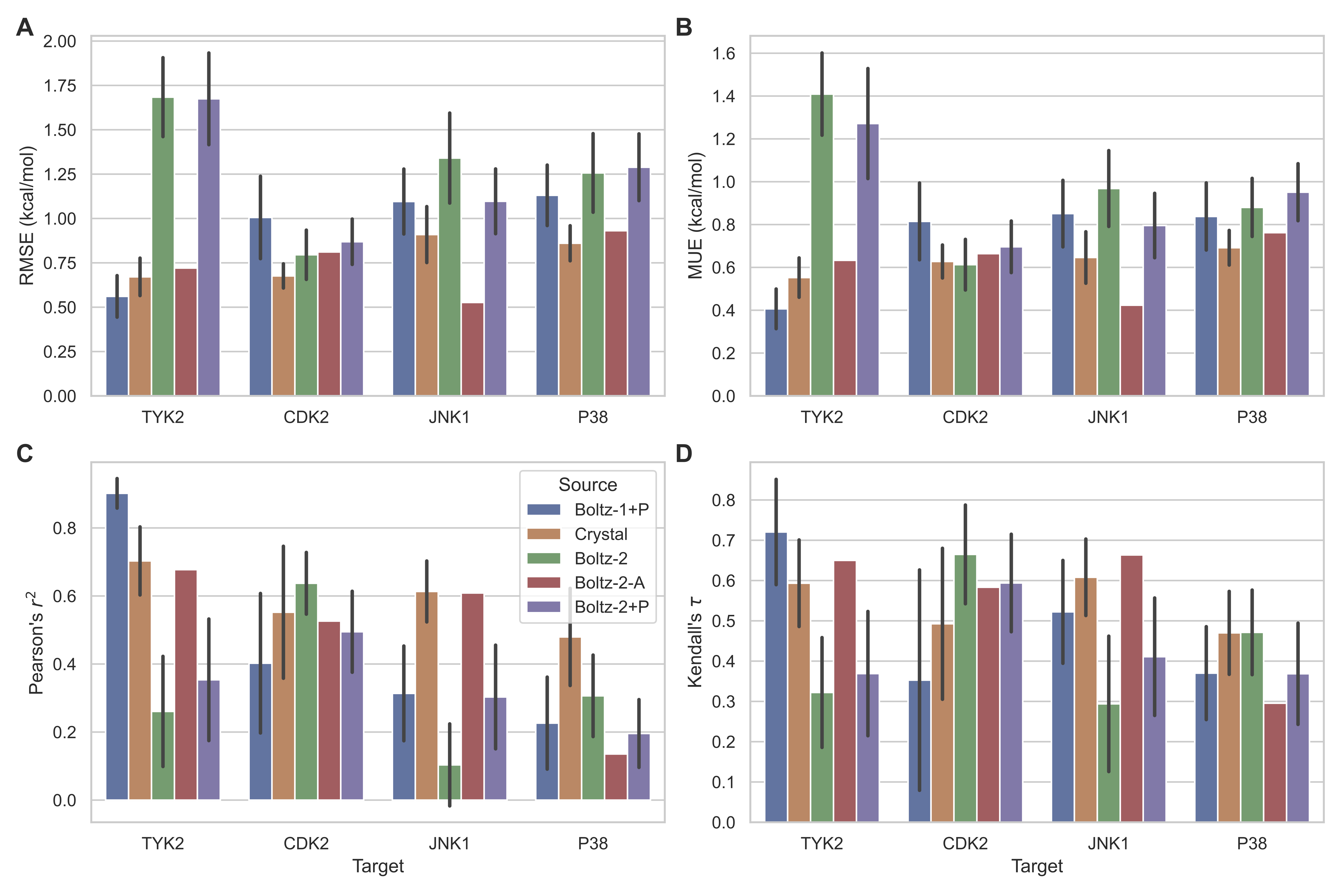}
    \caption{(A) Root mean square error (RMSE) between calculated and experimental $\Delta$G values for four target systems (TYK2, CDK2, JNK1, and P38), using input structures generated by the Boltz-1/2 models or taken from the protein-ligand benchmark. Three replicates were performed for all sets. RMSE error bars represent the standard deviation across replicates. (B) Mean unsigned error (MUE) between calculated and experimental $\Delta$G values. (C) Pearson’s $r^2$ between calculated and experimental $\Delta$G values. (D) Kendall’s $\tau$ correlation between calculated and experimental $\Delta$G values.}
    \label{fig:benchmark_4}
\end{figure}

We applied the Boltz-ABFE pipeline to 4 proteins from the FEP+ benchmark: CDK2, TYK2, JNK1, and P38. We evaluated the effect different structure prediction models had on the resultant ABFE predictions against experimental values. A summary of these results are shown in Figure \ref{fig:benchmark_4}. The different prediction models included: Boltz-2 without redocking (labeled "Boltz-2" in the Figure), Boltz-1 with redocking using POSIT ("Boltz-1+P"), and Boltz-2 with redocking using POSIT ("Boltz-2+P"). These results are compared against ABFE simulations starting from the ground-truth crystal structure with POSIT re-docking (labeled "crystal"). Additionally, the Figure includes results from the Boltz-2 Affinity module ("Boltz-2-A"). Table S2 of the supporting information contains a full list of these results with additional success metrics.

As Figure \ref{fig:benchmark_4} shows, the ABFE simulations starting from the crystal structure achieved the most consistent results over all the targets when considering all of the success metrics (RSME, MUE, $R^2$ and Kendall's $\tau$). However, starting from the crystal structure for this set does not guarantee the best result. For example, the best results for TYK2 start from Boltz1+P predicted structures.  In general, the ABFE results initiated from any of the Boltz predicted structures achieved satisfactory results with $\mathrm{MUE} < 1$ kcal/mol on average.
Surprisingly, the Boltz-1 with POSIT redocking (Boltz-1+P) performs better on average than both Boltz-2-based protocols, yielding a modest increase of 0.1 kcal/mol in mean unsigned error (MUE) compared to the crystal structure results on average. The poorer performance of Boltz-2 can be attributed to the TYK2 protein, where the predictions have $\mathrm{MUE}'s > 1$ kcal/mol. Upon closer inspection of the predicted TYK2 structures, we observed that Boltz-2 flipped a side-chain in the binding pocket compared to Boltz-1, which we presume to be the main reason for the difference in accuracy. Further work needs to be carried out to see how general this result might be. Finally, the top-down trained Boltz-2 Affinity module also performs well on this dataset, yielding correlation metrics that are slightly better than those from the Boltz-1+P ABFE simulations.



\section{Discussion and Conclusion}
In this work, we have investigated the quality of protein-ligand complex structures predicted by Boltz-1/2 for MD applications. 
With the exception of stereochemistry, the inference-time steering of Boltz-2 fixes major defects prevalent in Boltz-1-predicted structures, including steric clashes as well as incorrect aromaticity and bond order. In addition, we find that truncating low-confidence regions and including binding partners in the co-folding can be important for obtaining accurate structures.
Based on these findings, we have proposed Boltz-ABFE, a pipeline that corrects defects of predicted structures and allows to perform free energy simulations without requiring experimentally-determined protein-ligand complex structures.

We have demonstrated the effectiveness of our Boltz-ABFE pipeline based a set of 4 kinase targets.
While the inference-time steering of Boltz-2 proved effective in reducing the amount of structural errors, likely rendering the redocking step obsolete in future iterations of structure prediction models, this did not translate to significantly improved ABFE results. For example, a simple redocking step on top of Boltz-1 structures resulted in improved simulation results compared with the inference-time steered Boltz-2.
We note that the sensitivity of ABFE simulation estimates with respect to the initial structure may serve as an orthogonal benchmarking metric for protein-ligand co-folding models that does not rely on the availability of experimental crystal structures, but rather on more abundant affinity measurements.

Although we have analyzed binding affinity estimation performance based on a broad set of metrics, we note that correlation metrics such as Pearson's $R$ mainly measure performance in hit-to-lead and lead optimization settings to rank binders by their affinity. In early stage drug discovery, including hit identification and target deconvolution, separating binders from non-binders is typically sufficient. For this type of application, even the worst performing model of TYK2 ($\mathrm{MUE}=1.26$ kcal/mol), would probably have the resolution to successfully perform this task. 

We found the Boltz-2 Affinity module to be competitive with Boltz-ABFE on the evaluation set at hand. However, it is worth noting that unlike free energy simulations, the Boltz-2 Affinity module was exclusively trained on experimental binding affinity data (top-down). While the Boltz-2 Affinity module seems to shows strong performance on well-studied protein classes such as the kinases we investigated here, presumably due to their prevalence in public databases such as Chembl \cite{zdrazil2024chembl}, performance varies strongly across different targets, which we have shown in the Boltz-2 work for a set of 8 targets internal to Recursion \cite{passaro2025boltz}.
In contrast, we expect the performance of Boltz-ABFE to be more robust across different targets due to the physical foundation and bottom-up parametrization of ABFE simulations.

There are multiple avenues for future improvements of the Boltz-ABFE pipeline. First, considering co-factors as part of the co-folding prediction could extend the domain of applicability to cases where those co-factors are important for binding. For targets that are not as well characterized, models such as AlphaFill \cite{hekkelman2023alphafill} could be used to predict the existence of relevant co-factors in an initial step.
Second, the ABFE offset effect \cite{khalak2021alchemical, chen2023enhancing} is much more problematic in early-stage drug discovery applications, where absolute binding free energy values need to be compared across different protein holo structures (hit identification) or even across different proteins (target deconvolution).
To this end, selecting a free energy simulation protocol that explicitly considers the protein apo state could reduce this offset \cite{khalak2021alchemical}.  
By further optimizing the Boltz-ABFE pipeline, we envision that free energy simulations without experimental crystal structures can significantly accelerate early stage drug discovery via accurate, structure-based affinity estimation.

\section{Methods}

\subsection{Boltz Inference}
The Boltz-1/2 \cite{wohlwend2024boltz, passaro2025boltz} models predict protein-ligand complex structures based on the protein sequence(s) and the ligand SMILES. As an additional input, it requires a Multiple Sequence Alignment (MSA) for each sequence, which we compute via ColabFold \cite{mirdita2022colabfold} and MMseqs2 \cite{steinegger2017mmseqs2} with default parameters.
For each protein-ligand complex, we sample 10 structures, using 10 recycling steps and the default of 200 reverse diffusion steps. In the case of Boltz-2 (version 2.0.0), we enable the inference-time steering potentials to reduce the amount of steric clashes and ligand chemistry errors (Boltz-2x).

\subsection{ProMod3 Homology Modelling}
In an attempt to resolve clashes, before preparation, Boltz-1 models were put through a homology modelling pipeline using ProMod3, which is built on OpenStructure \cite{biasini2013} with an option to let ProMod3 decide the sidechain orientations rather than keep the ones in the original Boltz-1 model.

\subsection{Structure Preparation}
Boltz-1 models were prepared for docking and ABFE using the OpenEye Spruce Toolkit \cite{openeye}. This splits the system into components, places and optimizes proton positions, including searching ligand tautomer states, and assigns partial charges to the system. The N-terminus and C-terminus were capped with acetyl (ACE) and N-methyl (NME) groups respectively. A docking receptor object was added to each OpenEye Design Unit (DU) output file using default parameters for inner and outer contour shapes. 

Components of each generated OpenEye DU were preprocessed, converting protein residue and atoms names to AMBER format, before being parameterized using BioSimSpace \cite{BioSimSpace}. The protein and ligand components were parameterized with the AMBER ff14SB \cite{amber14sb} and GAFF2 \cite{gaff2} force fields respectively.

Re-docking with POSIT and Hybrid used the OpenEye Toolkits 2023.2.3\cite{openeye}. In each case, the Boltz-1/2 model OpenEye DU containing the cognate ligand was used as the docking receptor. An initial conformational ensemble was generated from the ligand input in SMILES format using OpenEye Omega, with pre-defined 'POSE' Omega Sampling options, which are optimal for pose prediction. The resultant ensemble was then passed to either the Hybrid or POSIT docking program (POSIT optimizes structures from the input ensemble internally), and only a single output pose was requested. Clashes between ligand and protein, including heavy atoms, were explicitly allowed, in order to maximize the likelihood of a successful pose being returned from an imperfect template co-folded ligand structure.

\subsection{ABFE protocol}

Ligand parameters were assigned using the OpenFF 2.2.1 \cite{openff211}. Partial charges for the ligands were derived using the AM1-BCC-ELF10 \cite{am1bccelf10} method as implemented in OpenEye software \cite{openeye}.
Solvation of the protein-ligand complexes involving neutral ligands was performed using the TIP3P water model within a rhombic dodecahedron simulation box. A shell distance of 0.5 nm was maintained between the solute and its periodic neighbor. The system was neutralized with NaCl at a concentration of 0.15 M. For complexes with charged ligands, solvation was carried out in a cubic box, maintaining a shell distance of 0.8 nm. To neutralize these systems, alchemical ion restraints were applied to ions positioned at the corners of the simulation box \cite{alchemicalion}. The unbound (free) ligand state was solvated in a cubic box with dimensions of 4 nm x 4 nm x 4 nm.

Absolute Binding Free Energy (ABFE) calculations were conducted following a previously established protocol \cite{ABFE_protocol}. The system was equilibrated using a well-established protocol \cite{roe2020protocol} with multiple rounds of energy minimization and short MD simulations with decreasing strength of positional restraints to relax the complex without deviating too far from the initial structure. The alchemical transformation involved the annihilation of the ligand. For the unbound ligand (free leg), this process was stratified into 48 intermediate windows. In the bound state, charged ligands were annihilated over 96 windows, while neutral ligands were annihilated over 64 windows. Molecular dynamics simulations for each alchemical window were conducted for 5 ns using GROMACS version 2024.5 \cite{GROMACS}. The resulting free energy data from all windows were processed using the alchemlyb \cite{alchemlyb} to compute the final binding free energy estimates.

%
%
%

\bibliography{biblio}
\end{document}


\maketitle              
%

\clearpage 


\begin{table}[h!]
    \centering
    \begin{tabular}{lcccc}
        \toprule
                                      & Crystal & Native  & POSIT   & HYBRID  \\
        \midrule
        RMSE (kcal/mol)               & 0.7±0.1 & 1.1±0.3 & 0.6±0.1 & 1.5±0.3 \\
        MUE (kcal/mol)                & 0.6±0.1 & 0.8±0.2 & 0.4±0.1 & 1.2±0.2 \\
        Pearson's $r^2$               & 0.7±0.1 & 0.6±0.1 & 0.89±0.04 & 0.2±0.2 \\
        Pearson's r                   & 0.9±0.1 & 0.8±0.1 & 0.95±0.02 & 0.5±0.2 \\
        Spearman's $\rho$             & 0.8±0.1 & 0.9±0.1 & 0.9±0.1 & 0.5±0.2 \\
        Kendall's $\tau$              & 0.6±0.1 & 0.8±0.1 & 0.7±0.1 & 0.3±0.2 \\
        \bottomrule
    \end{tabular}\\
    \caption{Comparison of Absolute Binding Free Energy (ABFE) prediction performance across various model refinement strategies. Values represent the mean of three replicates calculated for the crystal structure, native Boltz-1 output, Boltz-1 + POSIT re-docking, and Boltz-1 + HYBRID re-docking. Error bars indicate the bootstrap standard deviation.}
    \label{tab:TYK2_ABFE}
\end{table}

\begin{table}[ht!]

\caption{Comparison of prediction performance metrics across four protein targets for different structure generation and scoring methods. 'Crystal' refers to calculations initiated from experimental crystal structures. 'Boltz-1+P' and 'Boltz-2+P' use co-folded structures refined with POSIT redocking. 'Boltz-2' uses the raw co-folded structure, and 'Boltz-2-A' refers to a affinity prediction from the Boltz-2 Affinity module. The root-mean-squared error (RMSE) and mean unsigned error (MUE) were computed after centering $\Delta G$ predictions to the experimental mean to account for the difference between experimental IC50 and ABFE estimates (calculated Kd).}
\label{tab:full_benchmark_comparison}
\resizebox{0.7\textwidth}{!}{\begin{tabular}{lccccc}
\toprule
\textbf{Metric} & \textbf{Crystal} & \textbf{Boltz-1+P} & \textbf{Boltz-2} & \textbf{Boltz-2+P} & \textbf{Boltz-2-A} \\
\midrule
\multicolumn{6}{l}{\textbf{TYK2}} \\
\cmidrule(r){1-1}
RMSE (kcal/mol)   & 0.66 & 0.56 & 1.71 & 1.63 & 0.72 \\
MUE (kcal/mol)    & 0.55 & 0.42 & 1.44 & 1.26 & 0.63 \\
Pearson's $r^2$   & 0.72 & 0.89 & 0.28 & 0.33 & 0.68 \\
Pearson's r       & 0.85 & 0.95 & 0.53 & 0.58 & 0.82 \\
Spearman's $\rho$ & 0.82 & 0.90 & 0.48 & 0.57 & 0.81 \\
Kendall's $\tau$  & 0.62 & 0.73 & 0.33 & 0.40 & 0.65 \\
\midrule
\multicolumn{6}{l}{\textbf{CDK2}} \\
\cmidrule(r){1-1}
RMSE (kcal/mol)   & 0.68 & 1.00 & 0.77 & 0.88 & 0.81 \\
MUE (kcal/mol)    & 0.64 & 0.78 & 0.61 & 0.72 & 0.66 \\
Pearson's $r^2$   & 0.56 & 0.35 & 0.64 & 0.51 & 0.53 \\
Pearson's r       & 0.75 & 0.59 & 0.80 & 0.71 & 0.73 \\
Spearman's $\rho$ & 0.70 & 0.53 & 0.86 & 0.75 & 0.81 \\
Kendall's $\tau$  & 0.51 & 0.42 & 0.67 & 0.58 & 0.58 \\
\midrule
\multicolumn{6}{l}{\textbf{JNK1}} \\
\cmidrule(r){1-1}
RMSE (kcal/mol)   & 0.89 & 1.12 & 1.31 & 1.10 & 0.53 \\
MUE (kcal/mol)    & 0.64 & 0.83 & 0.96 & 0.83 & 0.42 \\
Pearson's $r^2$   & 0.60 & 0.32 & 0.12 & 0.28 & 0.61 \\
Pearson's r       & 0.77 & 0.56 & 0.34 & 0.53 & 0.78 \\
Spearman's $\rho$ & 0.81 & 0.71 & 0.45 & 0.59 & 0.83 \\
Kendall's $\tau$  & 0.61 & 0.54 & 0.28 & 0.42 & 0.66 \\
\midrule
\multicolumn{6}{l}{\textbf{P38}} \\
\cmidrule(r){1-1}
RMSE (kcal/mol)   & 0.86 & 1.16 & 1.25 & 1.28 & 0.93 \\
MUE (kcal/mol)    & 0.69 & 0.81 & 0.90 & 0.94 & 0.76 \\
Pearson's $r^2$   & 0.48 & 0.22 & 0.33 & 0.21 & 0.14 \\
Pearson's r       & 0.69 & 0.47 & 0.57 & 0.46 & 0.34 \\
Spearman's $\rho$ & 0.66 & 0.50 & 0.67 & 0.52 & 0.45 \\
Kendall's $\tau$  & 0.46 & 0.36 & 0.49 & 0.37 & 0.30 \\
\midrule
\multicolumn{6}{l}{\textbf{Overall}} \\
\cmidrule(r){1-1}
RMSE (kcal/mol)   & 0.80 & 1.01 & 1.26 & 1.23 & 0.77 \\
MUE (kcal/mol)    & 0.64 & 0.74 & 0.96 & 0.93 & 0.64 \\
Pearson's $r^2$   & 0.57 & 0.39 & 0.33 & 0.30 & 0.42 \\
Pearson's r       & 0.75 & 0.60 & 0.55 & 0.54 & 0.61 \\
Spearman's $\rho$ & 0.73 & 0.63 & 0.62 & 0.59 & 0.67 \\
Kendall's $\tau$  & 0.53 & 0.48 & 0.44 & 0.43 & 0.50 \\
\bottomrule
\end{tabular}}
\end{table}

\begin{figure}
    \centering
    \includegraphics[width=0.9\linewidth]{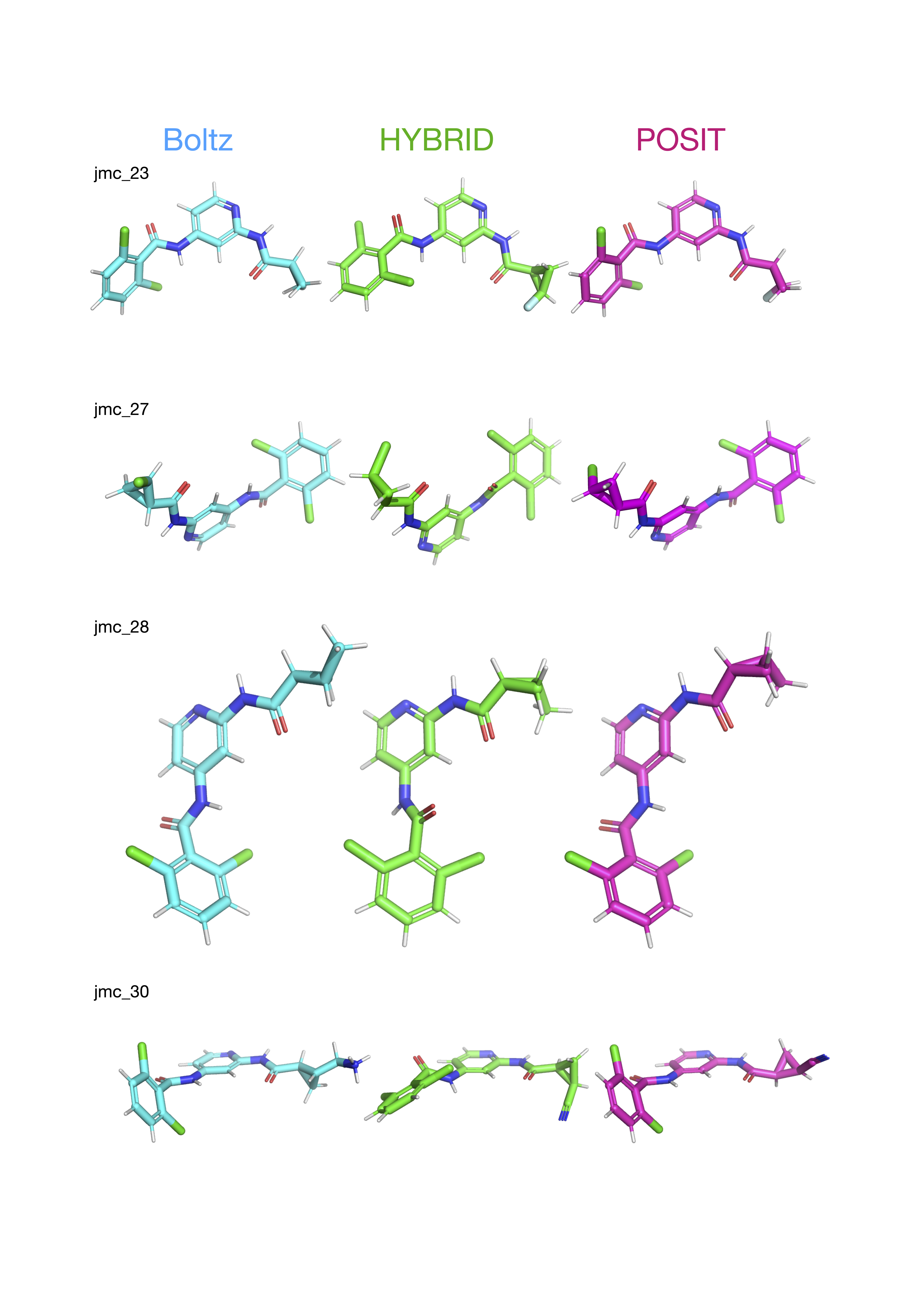}
    \caption{Overlay of the four ligands that Boltz-1 generates model with incorrect chemistry in the TYK2 set}
    \label{fig:overlau}
\end{figure}

\begin{figure}

    \includegraphics[width=0.9\linewidth]{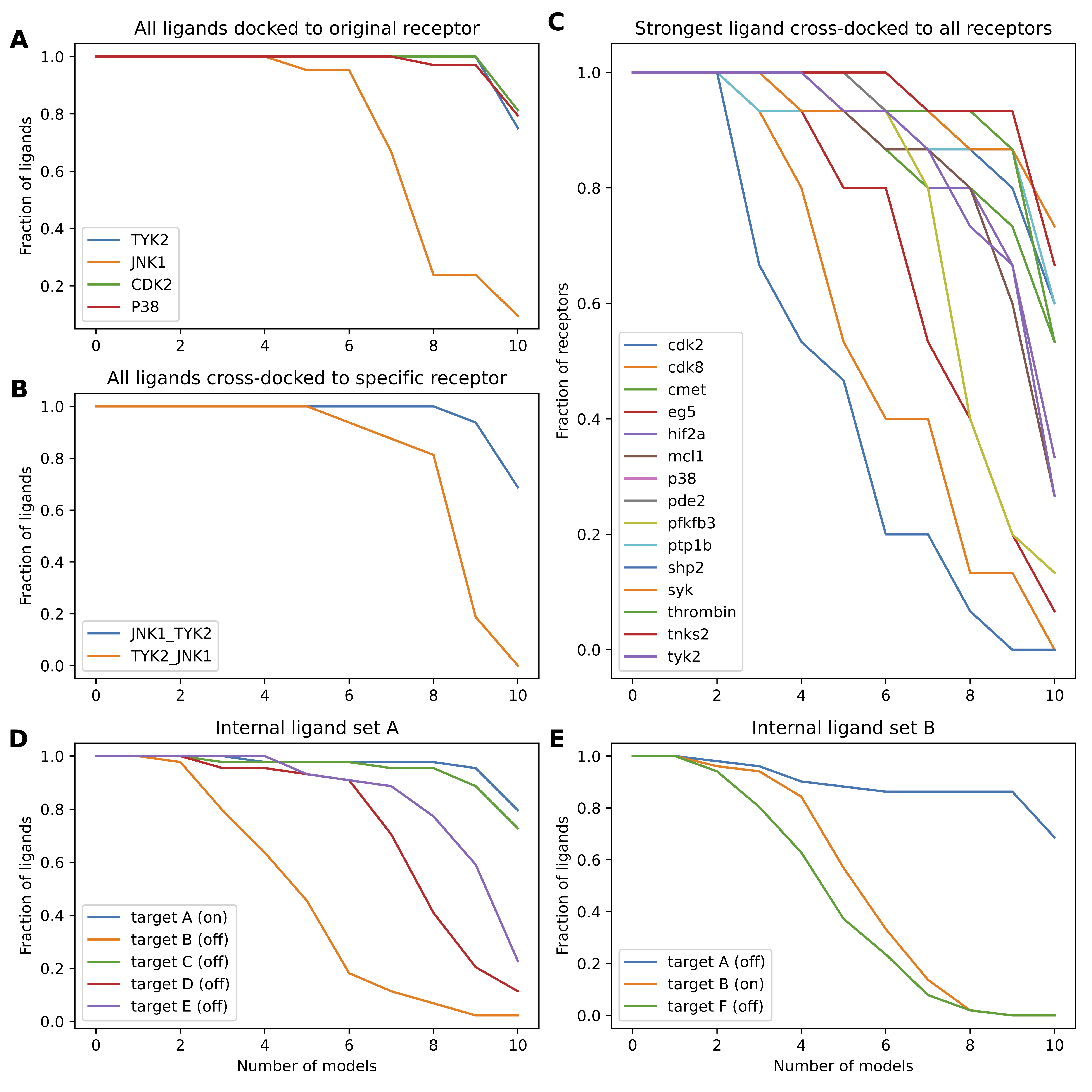}
    \caption{Success rates for generating clash-free structures with the Boltz-1 model across several co-folding scenarios. In general, generating 10 models per pair proved sufficient to yield at least one structure compatible with the OpenEye Spruce pipeline for most systems tested. The y-axis represents the percentage of ligands that yield at least X successful models (x-axis). The panels depict several co-folding scenarios: (A) Public ligands (TYK2, CDK2, JNK1, and P38) docked to their cognate receptors. (B) Public ligands cross-docked to a different receptor (TYK2 ligands to JNK1 and vice versa). Success rates varied significantly by system; TYK2 exhibited high processing success rates for both native and cross-docked ligands, whereas JNK1 displayed lower rates. (C) The highest-affinity ligand from each of the 15 OpenFF benchmark systems cross-docked to the remaining targets. (D, E) Two internal ligand sets docked to their on-targets and related off-targets. Among these internal sets, target A showed high success rates for both binders and non-binders, while target B yielded lower rates for both categories. The results were inconclusive as to whether Boltz-1 models were more likely to be successfully prepared for known binders versus non-binders.}
    \label{fig:success_rate_1}
\end{figure}

\begin{figure}
    \centering
    \includegraphics[width=1.0\linewidth]{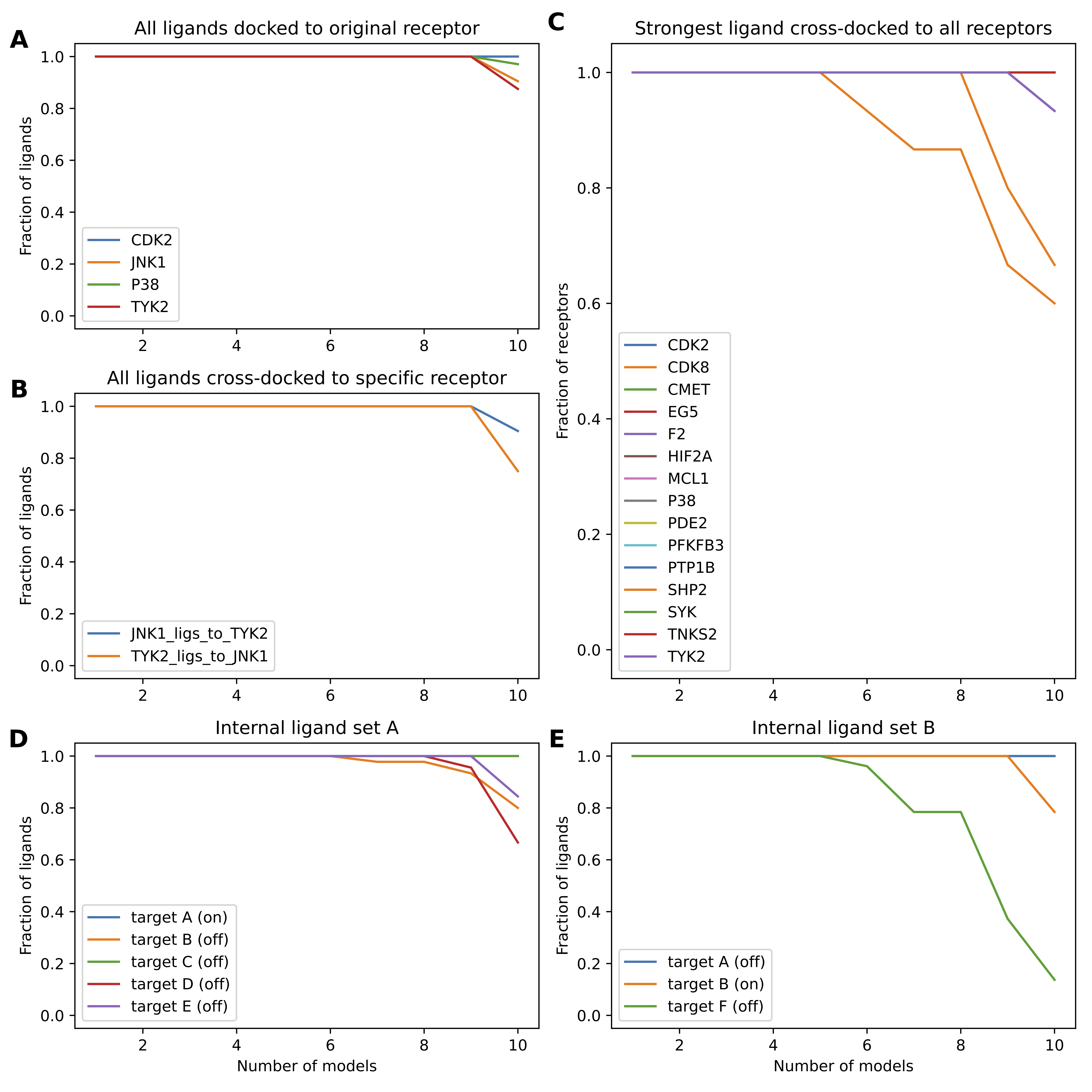}
    \caption{Success rates for generating clash-free structures using the Boltz-2 model across several co-folding scenarios. These results show a consistent and significant improvement over the Boltz-1 model. Notably, Boltz-2 enhances success rates even for systems that were particularly challenging for Boltz-1. For instance, while the CDK8 and SHP2 targets still exhibit relatively low success rates, they are markedly improved compared to their Boltz-1 results (Panel C). Similarly, internal target 7, which performed poorly with Boltz-1, shows a substantial increase in its success rate with Boltz-2 (Panel E).}
    \label{fig:success_rate_2}
\end{figure}